\documentclass[aps,pre,floats,twocolumn,superscriptaddress]{revtex4}
\usepackage{amssymb}

\usepackage{graphicx}
\usepackage{subfigure}
\usepackage[tbtags]{amsmath}
\usepackage{fancybox}
\usepackage{array}
\usepackage{color}
\usepackage{dcolumn}
\usepackage{amsmath}

\begin{document}
\title{Order parameter analysis for low-dimensional behaviors of coupled phase-oscillators}
\author{Jian Gao}
\affiliation{Department of Physics and the Beijing-Hong Kong-Singapore Joint Centre for Nonlinear and Complex Systems (Beijing), Beijing Normal University, Beijing 100875, China}
\author{Can Xu}
\affiliation{Department of Physics and the Beijing-Hong Kong-Singapore Joint Centre for Nonlinear and Complex Systems (Beijing), Beijing Normal University, Beijing
100875, China}
\author{Yuting Sun}
\affiliation{Department of Physics and the Beijing-Hong Kong-Singapore Joint Centre for Nonlinear and Complex Systems (Beijing), Beijing Normal University, Beijing 100875, China}
\author{Zhigang Zheng}
\email[]{zgzheng@bnu.edu.cn}
\affiliation{Department of Physics and the Beijing-Hong Kong-Singapore Joint Centre for Nonlinear and Complex Systems (Beijing), Beijing Normal University, Beijing 100875, China}
\affiliation{College of Information Science and Engineering, Huaqiao University, Xiamen 361021, China}
\date{\today}
\begin{abstract}
Coupled phase-oscillators are important models related to synchronization. Recently, Ott-Antonsen(OA) ansatz is developed and used to get low-dimensional collective behaviors in coupled oscillator systems. In this paper, we develop a simple and concise approach based on the equations of order parameters, namely, order parameter analysis, with which we point out that the OA ansatz is rooted in the dynamical symmetry of the order parameters. With our approach the scope of the OA ansatz is identified as two conditions, i.e., infinite size of the system and only three nonzero Fourier coefficients of the coupling function. Coinciding with each of the conditions, a distinctive system out of the scope is taken into account and discussed with the order parameter analysis. Two approximation methods are introduced respectively, namely the ensemble approach and the dominating-term assumption.
\end{abstract}
\maketitle
\section{Introduction}
Understanding the intrinsic mechanism of collective behaviors of coupled units has become a focus for a variety of fields, such as biological neurons, circadian rhythm, chemically reacting cells, and even social systems \cite{kuramoto1984chemical,acebron2005kuramoto,strogatz2000kuramoto,pikovsky2002synchronization,dorogovtsev2008critical,
arenas2008synchronization,zheng1998phase}. Some properties of collective behaviors depend on the complexity of the system, while the other properties, such as phase transition, may be described by low-dimensional dynamics with macroscopic variables \cite{watanabe1993integrability,watanabe1994constants,marvel2009identical,marvel2009invariant,ott2008low}. Discovering the method to simplify the system is just as important and as fascinating as the discovery of the complexity of it.

Like most cases in physics, simplification and low-dimensional reduction are associated with some symmetry of the system. As the identity of gas particles is the foundation of statistical mechanics and collective variables as temperature and pressure, the identity of the coupled units in a complex system is also related to some order parameters. In previous works, in the limit of large number of oscillators $N\rightarrow\infty$ with special coupling function this work has been done by  Ott-Antonsen(OA) ansatz \cite{ott2008low} for oscillators with nonidentical parameters. As for the oscillators with identical parameters the same result was also got from group theory analysis in \cite{marvel2009identical} called Watanabe-Strogatz's approach.

In this paper, we will show a different way in which the low-dimensional reduction is a natural consequence of the symmetry of the system by taking the order parameters as collective variables, namely the order parameter analysis. Our approach is simple and concise for oscillators with both identical and nonidentical parameters. We can also get the scope of the OA ansatz. Two more cases beyond the scope are further discussed with appropriate approximations. With our approach we show that the OA ansatz can be used beyond its scope with the approximations works.

The model discussed in this paper is the all-connected coupled phase-oscillators. In the first section, the dynamical equations for order parameters are derived. The OA ansatz is got naturally from the symmetry of the order parameter equations, with its scope as the limit of infinitely many oscillators and the condition that only three Fourier coefficients of the coupling function are nonzero. In the second and third sections, we will consider two approximate use of this ansatz beyond its scope, the case of a finite-size system and the case of coupled oscillator systems with more complicated coupling functions, with approximations respectively, i.e., the ensemble approach and the dominating-term assumption.

\section{Equations for order parameters}
The famous Kuramoto model for the process of synchronization attracts much attentions upon it is proposed and has been developed for decades. This model consists of a population of N coupled phase oscillators $\{\varphi_{j}\}$ with natural frequencies $\{\omega_{j}\}$ , and the dynamics are governed by
\begin{equation}\label{e1}
\dot\varphi_{j}(t)=\omega_{j}-\sum_{k=1}^{N}A_{jk}\sin(\varphi_{k}-\varphi_{j}),\ \ \ j=1,\dots,N.
\end{equation}
With the mean-field coupling $A_{jk}=K/N$ and the definition of order parameter
\begin{equation}
\alpha=\frac{1}{N}\sum_{j=1}^{N}e^{\mathrm{i}\varphi_{j}},
\end{equation}
the equations Eq. (\ref{e1}) can be rewritten as
\begin{equation}
\dot\varphi_{j}(t)=\omega_{j}-\frac{K}{2i}(\alpha e^{-\mathrm{i}\varphi_{j}}-\bar\alpha e^{\mathrm{i}\varphi_{j}}),\ \ \ j=1,\dots,N,
\end{equation}
where $\mathrm{i}$ is the imaginary unit, $\bar\alpha$ is the complex conjugate of $\alpha$. Apart form parameters $K$ and $\omega_{j}$, the dynamics of each phase variable $\varphi_{j}$ depends only on itself and $\alpha$. It is the important character of this mean-field model, and the order parameter $\alpha$ is always used to describe the state of the system, as the system is in synchronous states if and only if $|\alpha|=1$.

A more general form of the mean-field model with phase oscillators can be written as
\begin{equation}\label{ew1}
\dot\varphi_{j}(t)=F(\boldsymbol{\alpha},\varphi_{j},\boldsymbol{\beta},\boldsymbol{\gamma_{j}}),\ \ j=1,\dots,N,
\end{equation}
where $F(\boldsymbol{\alpha},\varphi_{j},\boldsymbol{\beta},\boldsymbol{\gamma_{j}})$ is any smooth, real, $2\pi$-periodic function for $\varphi_{j}$. $\boldsymbol{\alpha}=(...\alpha_{-1},\alpha_{0},\alpha_{1}...)$  is the order parameter with the $n$th order parameter $\alpha_{n}$ defined as
\begin{equation}
\alpha_{n}=\frac{1}{N}\sum_{j=1}^{N}e^{\mathrm{i}n\varphi_{j}}.
\end{equation}
$\boldsymbol{\beta}=(\beta_{1},\beta_{2}...\beta_{s})$ is the identical parameter such as the coupling strength $K$ which is identical for all the oscillators. $\boldsymbol{\gamma_{j}}=(\gamma_{j}^{(1)},\gamma_{j}^{(2)}...\gamma_{j}^{(l)})$ is the nonidentical parameter such as the natural-frequency $\{\omega_{j}\}$ which is nonidentical for the oscillators and usually has a distribution among the system.

Almost all the mean-field models based on the Kuramoto model belong to this general category. In the following, we will build our approach for this general model to explore the conditions in which we can get the low-dimensional description of the system Eq. (\ref{ew1}).

\subsection{Identical oscillators}
To begin with, let us consider the simple case of identical oscillators, e.g., $\omega_{j}=\omega^{*}$ for $j=1,2,\dots,N$ in the Kuramoto model, which reads
\begin{equation}\label{e2}
\dot\varphi_{j}(t)=F(\boldsymbol{\alpha},\varphi_{j},\boldsymbol{\beta}),j=1,\dots,N.
\end{equation}
In the limit of infinitely many oscillators $N\rightarrow\infty$, let $\rho(t,\varphi)d\varphi$ denote the fraction of oscillators that lie between $\varphi$ and $\varphi+d\varphi$. Because each oscillator in Eq.(\ref{e2}) moves with the angular velocity $v_{j}=F(\boldsymbol{\alpha},\varphi_{j},\boldsymbol{\beta})$, the single oscillator density $\rho(t,\varphi)$ obeys the continuity equation as
\begin{equation}\label{ea1}
\begin{aligned}
&\frac{\partial\rho}{\partial t}+\frac{\partial(\rho v)}{\partial\varphi}=0,\\
&v=F(\boldsymbol{\alpha},\varphi,\boldsymbol{\beta}).
\end{aligned}
\end{equation}

If the phase density $\rho(t,\varphi)$ is known, all of macroscopic properties of the system can be got through some statistical average, such as order parameters $\alpha_{n}$ as
\begin{equation}\label{e3}
\alpha_{n}=\int_{0}^{2\pi}\rho(t,\varphi)e^{\mathrm{i}n\varphi}d\varphi, n=1,2,\dots.
\end{equation}
If we are only concerned with the collective or macroscopic state of the system, as in this paper, the macroscopic description of the phase density is equivalent to the microscopic description of phases of all the oscillators.

Moreover, for the coupling function $F(\boldsymbol{\alpha},\varphi,\boldsymbol{\beta})$ is always $2\pi$-periodic for $\varphi$, we have the Fourier expansion of $F(\boldsymbol{\alpha},\varphi,\boldsymbol{\beta})$,
\begin{equation}
F(\boldsymbol{\alpha},\varphi,\boldsymbol{\beta})=\sum_{j=-\infty}^{\infty}f_{j}(\boldsymbol{\alpha},\boldsymbol{\beta})e^{\mathrm{i}j\varphi}.
\end{equation}

With the the definition Eq. (\ref{e3}), the dynamics of order parameters can be got as
\begin{equation}
\begin{aligned}
\dot\alpha_{n}&=\int_{0}^{2\pi}ine^{\mathrm{i}n\varphi}\dot\varphi d\varphi\\
              &=\int_{0}^{2\pi}ine^{\mathrm{i}n\varphi}F(\boldsymbol{\alpha},\varphi,\boldsymbol{\beta})d\varphi,
\end{aligned}
\end{equation}
where $n=1,2,\dots$. Substituting the expansion of $F(\boldsymbol{\alpha},\varphi,\boldsymbol{\beta})$ into these equations, we have the closed equations for the dynamics of order parameters as
\begin{equation}\label{e4}
\dot\alpha_{n}=\mathrm{i}n\sum_{j=-\infty}^{\infty}f_{j}(\boldsymbol{\alpha},\boldsymbol{\beta})\alpha_{j+n}, n=1,2,\dots.
\end{equation}

On the other hand, from Eq. (\ref{e3}) the order parameters are exactly Fourier components of the phase density. If all the order parameters are known, we have
\begin{equation}\label{e5}
\rho(t,\varphi)=\frac{1}{2\pi}\sum_{j=-\infty}^{\infty}\alpha_{j}e^{-\mathrm{i}j\varphi}.
\end{equation}
Hence if we know all the initial values of order parameters, with the dynamical equations Eq. (\ref{e4}) and the Fourier transformation Eq. (\ref{e5}), the system is identified, which is the same as the dynamical equations Eq. (\ref{e2}) for phase variables $\{\varphi_{j}\}$ and the continuity equation Eq. (\ref{ea1}) for the phase density $\rho(t,\varphi)$. Therefore, we can choose either the order parameters, the phase variables or the phase density to perform the analysis of the collective behaviors of the coupled phase oscillators.

These three descriptions of the system, i.e. phase variables, density of phase and order parameters, correspond to the dynamical, statistical and macroscopic descriptions of the system, respectively, and they are equivalent to each other in the limit of infinitely many oscillators  $N\rightarrow\infty$. As a matter of fact, the Ott-Antonsen ansatz \cite{ott2008low} is based on the representation of the density of phase, and the Watanabe-Strogatz's approach \cite{marvel2009identical} is based on the scenario of the phase variables. In the following, we will take our approach on the base of order parameters.

The complexity of dynamical equations for order parameters Eq. (\ref{e4}) depends on the coupling function, or explicitly, on the Fourier expansion of $F$. First, let us consider the simplest case, with only the first three nonzero terms of the Fourier expansion,
\begin{equation}\label{e6}
F(\boldsymbol{\alpha},\varphi)=f_{1}(\boldsymbol{\alpha})e^{\mathrm{i}\varphi}+f_{-1}(\boldsymbol{\alpha})e^{-\mathrm{i}\varphi}+f_{0}(\boldsymbol{\alpha}).
\end{equation}
Because $F$ is a real function, we have $f_{-1}=\bar f_{1}$ where $\bar f_{1}$ means the complex conjugate of $f_{1}$. With Eq. (\ref{e6}), the dynamics of order parameters Eq. (\ref{e4}) becomes
\begin{equation}\label{e7}
\dot\alpha_{n}=\mathrm{i}n(f_{1}\alpha_{n+1}+\bar f_{1}\alpha_{n-1}+f_{0}\alpha_{n}),
\end{equation}
with $n\geq0$ and $\alpha_{-n}=\bar\alpha_{n}$. The recursion form of these equations shows that there is some structure for order parameters with which we can simplify the system and get some low-dimensional equations for the high-dimensional coupled oscillator dynamics of the system. In fact, by choosing an invariant manifold as $\alpha_{n}=\alpha_{1}^{n},n\geq 0$, all the equations for $\alpha_{n}$ is reduced to
\begin{equation}\label{ew6}
\dot\alpha_{1}=\mathrm{i}(f_{1}(\alpha_{1})\alpha_{1}^{2}+\bar f_{1}(\alpha_{1})+f_{0}(\alpha_{1})\alpha_{1})
\end{equation}
for $n=1,2,\dots$. The infinitely many dynamical equations of order parameters is thus reduced to a single equation on this manifold and the corresponding phase density defined by Eq. (\ref{e5}) is the so-called Poisson kernel distribution as
\begin{equation}
\rho(t,\varphi)=\frac{1}{2\pi}\frac{1-r^{2}}{1-2r\cos(\varphi-\phi)+r^{2}},
\end{equation}
where $\alpha_{1}=re^{\mathrm{i}\phi}$. If we choose the initial state on this manifold, the state will never evolve out of it, which is called the invariant manifold of the dynamical system Eq. (\ref{e7}). This is exactly the low-dimensional behavior of the system we are looking for. With our approach, the manifold is derived naturally and concisely. We will call this manifold the Poisson manifold in this paper as in \cite{marvel2009identical} where the dynamics of $\alpha_{1}$, as Eq. (\ref{ew6}), is got from the group theory analysis for Josephson junction arrays.

\subsection{Nonidentical parameters}
Furthermore, the low-dimensional behavior is not confined to the system of identical oscillators. The order parameter analysis we used above can also be used in a more general case of nonidentical oscillators with nonidentical parameters.

Firstly, let us consider a system of nonidentical oscillators with a discrete distribution of the parameter $\boldsymbol{\omega}=\{\omega^{(1)},\omega^{(2)},...\omega^{(p)}\}$. This distribution naturally separates the oscillators into $p$ groups, and each group has the same parameter $\omega^{(p)}$. In this case, the phase variable for oscillators is denoted by $\varphi_{j}^{(r)}$ which means the $j$th oscillator with the same parameter $\omega^{(r)}$.

For each group we can define a local order parameter as
\begin{equation}
\alpha_{n}^{(r)}=\frac{1}{n^{r}}\sum_{l=1}^{n^{r}}e^{\mathrm{i}n\varphi_{l}^{(r)}},n=1,2,\dots,
\end{equation}
where $n^{r}$ is the number of oscillators in the group with $\omega^{(r)}$. The order parameter of the system denoted by $\alpha_{n}$ is defined by
\begin{equation}
\alpha_{n}=\sum_{r=1}^{p}\alpha_{n}^{(r)}\frac{n_{r}}{n},n=1,2,\dots.
\end{equation}
Assume that the coupling function for each oscillator $F(\varphi_{j})$ depends only on the order parameters whether for the whole system or the groups, then equations for phase oscillators are
\begin{equation}
\dot\varphi_{j}^{(r)}(t)=F(\boldsymbol{\alpha},\boldsymbol{\alpha^{(r)}},\varphi_{j}^{(r)},\omega^{(r)},\boldsymbol{\beta}),j=1,\dots,n^{r},
\end{equation}
where $r=1,2\ldots p$. $\boldsymbol{\alpha^{(r)}}=(...\alpha_{-1}^{(r)},\alpha_{0}^{(r)},\alpha_{1}^{(r)}...)$ and  $\boldsymbol{\alpha}=(...\alpha_{-1},\alpha_{0},\alpha_{1}...)$ are the local and global order parameters, respectively.

As the coupling function $F(\boldsymbol{\alpha},\boldsymbol{\beta},\boldsymbol{\alpha^{(r)}},\omega^{(r)},\varphi^{(r)})$ is $2\pi$-periodic for $\varphi_{j}^{(r)}$, by performing Fourier expansion of $F$, the equations for order parameters of each group read
\begin{equation}\label{e8}
\dot\alpha_{n}^{(r)}=\mathrm{i}n\sum_{j=-\infty}^{\infty}f_{j}(\boldsymbol{\alpha},\boldsymbol{\beta},\boldsymbol{\alpha^{(r)}},\omega^{(r)})\alpha_{j+n}^{(r)},
\end{equation}
where $r=1,2\ldots p$, $n=1,2,\dots$ and $F(\varphi^{(r)})=\sum_{j=-\infty}^{\infty}f_{j}e^{\mathrm{i}j\varphi^{(r)}}$.

The equations Eq. (\ref{e8}) are closed for local order parameters, and the local order parameters here can be regarded as the coordinates of the system which are equivalent to phase variables. Each group is almost separated from others and the only relation for the groups is the dependence of the coupling function on order parameters of the system.

In the simplest case when only the first three terms of Fourier expansion are nonzero, we have
\begin{equation}
\dot\alpha_{n}^{(r)}=\mathrm{i}n(f_{1}\alpha_{n+1}^{(r)}+\bar f_{1}\alpha_{n-1}^{(r)}+f_{0}\alpha_{n}^{(r)}),
\end{equation}
with $r=1,2\ldots p$ and $n=1,2,\dots$. For each group, $\alpha_{n}^{(r)}=(\alpha_{1}^{(r)})^{n}$ is obviously a solution for the equations which reduce all the equations for $\alpha_{n}^{(r)}$ to the same one as
\begin{equation}\label{e9}
\dot\alpha_{1}^{(r)}=\mathrm{i}(f_{1}(\alpha_{1}^{(r)})^{2}+\bar f_{1}+f_{0}\alpha_{1}^{(r)}),
\end{equation}
where $r=1,2\ldots p$. The Poisson manifold is an invariant manifold for each group, and the order parameters of the system read
\begin{equation}
\alpha_{n}=\sum_{r=1}^{p}(\alpha_{1}^{(r)})^{n}\frac{n_{r}}{n}, n=1,2,\dots,
\end{equation}
which is determined by $\alpha_{1}^{(r)},r=1,2\ldots p$. Therefore, the Fourier coefficients of the coupling function $f_{j}(\boldsymbol{\alpha},\boldsymbol{\beta},\boldsymbol{\alpha^{(r)}},\omega^{(r)})$ depends on $\alpha_{1}^{(r)},r=1,2\ldots p$ only. The system described by the local order parameters is governed by a group of Poisson manifolds with the low-dimensional dynamics Eq. (\ref{e9}).

In the limit with $n^{r}\rightarrow\infty,r=1,2...p$ and $p\rightarrow\infty$, the distribution of parameter $\omega$ as $\{n^{1},n^{2},...n^{p}\}$ for $\{\omega^{(1)},\omega^{(2)},...\omega^{(p)}\}$ has the continuous form denoted by the function $g(\omega)$ and the local order parameters $\alpha_{1}^{(r)},r=1,2\ldots p$ become the function of $\omega$, i.e., $\alpha_{1}(\omega)$. Replacing the summation by the integral over $\omega$, we get the continuous form of the dynamical equations as
\begin{equation}\label{e10}
\begin{aligned}
\dot\alpha_{1}(\omega)=&\mathrm{i}(f_{1}(\omega,\alpha_{1}(\omega),\boldsymbol{\alpha^{*}})(\alpha_{1}(\omega))^{2}+\bar f_{1}(\omega,\alpha_{1}(\omega),\boldsymbol{\alpha^{*}})\\
&+f_{0}(\omega,\alpha_{1}(\omega),\boldsymbol{\alpha^{*}})\alpha_{1}(\omega)),
\end{aligned}
\end{equation}
where $\boldsymbol{\alpha^{*}}=(...\alpha_{-1}^{*},\alpha_{0}^{*},\alpha_{1}^{*}...)$ are functionals of $\alpha_{1}(\omega)$ as
\begin{equation}\label{e11}
\alpha_{n}^{*}=\int(\alpha_{1}(\omega))^{n}g(\omega)d\omega, n=1,2,\dots.
\end{equation}

Even though for each specific $\omega$ the approach has already reduced the dynamics of the oscillators to the dynamic of $\alpha_{1}(\omega)$ as Eq. (\ref{e10}), it is still hard to get any analytical results as Eq. (\ref{e10}) depends on the integral Eq. (\ref{e11}) which cannot be expressed by functions of $\alpha_{1}(\omega)$. On the other hand, for some specific choice of $g(\omega)$, the integral Eq. (\ref{e11}) can be obtained analytically, in which case we can get the simpler form of Eq. (\ref{e10}).

What we are looking for is a two-step reduction. The first step is finished by introducing the Poisson manifold with which we have reduced the dynamic of phase oscillators to the dynamic of a group of Poisson manifolds, i.e., Eq. (\ref{e10}). The second step is rooted in the relation between the Poisson manifolds in the group which will reduce the dynamics of the group further to the dynamic of a specific Poisson manifold in this group.

For instance, in the case that the function $\alpha_{1}(\omega)$ is analytical which can be extended to the complex plane, and the distribution of $\omega$ is the Lorentzain distribution as
\begin{equation}
g(\omega)=\frac{\Delta}{\pi((\omega-\omega_{0})^{2}+\Delta^{2})},
\end{equation}
the integral Eq. (\ref{e11}) can be obtained by residue theorem as $\alpha_{n}^{*}=(\alpha_{n}(-\mathrm{i}))^{n}$. Setting $\omega=-\mathrm{i}$, Eq. (\ref{e10}) becomes
\begin{equation}\label{e12}
\begin{aligned}
\dot\alpha_{1}(-\mathrm{i})=&i(f_{1}(-\mathrm{i},\alpha_{1}(-\mathrm{i}))(\alpha_{1}(-\mathrm{i}))^{2}+\bar f_{1}(-\mathrm{i},\alpha_{1}(-\mathrm{i}))\\
&+f_{0}(-\mathrm{i},\alpha_{1}(-\mathrm{i}))\alpha_{1}(-\mathrm{i})),
\end{aligned}
\end{equation}
which is closed for order parameter $\alpha_{1}(-\mathrm{i})=\alpha_{1}^{*}$. Then, the behavior of $\alpha_{1}^{*}$, together with the collective behaviors of the system, is determined by Eq. (\ref{e12}).

Another solvable example is the case when the distribution of natural frequencies of oscillators is the Dirac's delta distribution function $g(\omega)=\delta(\omega-\omega_{0})$. The integral Eq. (\ref{e11}) can be worked out as $\alpha_{n}^{*}=(\alpha_{1}(\omega_{0}))^{n}$. Setting $\omega=\omega_{0}$ in Eq. (\ref{e10}), the model is reduced to the network of identical oscillators which we discussed above.

The approach shown above indicates there are two steps of the reduction scheme. The first step is the reduction from the equation Eq. (\ref{e8}) to Eq. (\ref{e9}) or the continuous form Eq. (\ref{e10}), which means the choice of the Poisson manifold of the system. This is related to the the symmetry of the system or the recursion form of order parameters equations. The second step is the further operation from Eq. (\ref{e10}) to Eq. (\ref{e12}), which depends on the special choice of distribution of nonidentical parameters. This reduction with the Lorentzain distribution was first found in \cite{ott2008low} with an ansatz for low-dimensional manifold, namely the OA ansatz. For the case of the Dirac's delta distribution function, this reduction was discussed comprehensively in \cite{marvel2009identical} with group theory analysis of the system. In our approach, the reduction comes from the same basis, i.e., the symmetry of order parameter equations.

Along the approach, we can also consider more nonidentical parameters, such as the location of oscillators $x$ for example in \cite{Laing2009chimera}, where the coupling function depends on the locations $\boldsymbol{x}$ as $F(\boldsymbol{\alpha},\boldsymbol{\alpha(\omega,x)},\varphi,\omega,\boldsymbol{x},\boldsymbol{\beta})$ with the distributions $g(\omega)$ and $q(x)$. The order parameter $\boldsymbol{\alpha}$ of the system is defined as
\begin{equation}\label{e13}
\alpha_{n}=\int\int\alpha_{n}(\omega,x)g(\omega)q(x)d\omega dx,n=1,2,\dots.
\end{equation}
When the Fourier expansion of the coupling function for $\varphi$ contains only the first three terms, for each specific $\omega$ and $x$, the relation $\alpha_{n}(\omega,x)=(\alpha_{1}(\omega,x))^{n}$ is exactly the solution for dynamical equations, which reads
\begin{equation}\label{e14}
\dot\alpha_{n}(\omega,x)=\mathrm{i}n(f_{1}\alpha_{n+1}(\omega,x)+\bar f_{1}\alpha_{n-1}(\omega,x)+f_{0}\alpha_{n}^{*}(\omega,x)).
\end{equation}

If $\alpha(\omega,x)$ is analytic for $\omega$ and $g(\omega)$ is the Lorentzain distribution, the integral Eq. (\ref{e13}) can be obtained for $\omega$ as
\begin{equation}
\begin{aligned}
\alpha_{n}&=\int\int\alpha_{n}(\omega,x)g(\omega)q(x)d\omega dx\\
&=\int\alpha_{n}(-\mathrm{i},x)q(x)dx,
\end{aligned}
\end{equation}
where $n=1,2,\dots$. Setting $\gamma_{n}(x)\equiv\alpha_{n}(-\mathrm{i},x)=(\alpha_{1}(-\mathrm{i},x))^{n}$ and $\omega=-\mathrm{i}$ in the dynamical equation Eq. (\ref{e14}), we have
\begin{equation}\label{e15}
\begin{aligned}
\dot\gamma_{1}(x)=&\mathrm{i}(f_{1}(\omega,\gamma_{1}(x),\boldsymbol{\alpha})\gamma_{1}(x)^{2}\\
&+\bar f_{1}(\omega,\gamma_{1}(x),\boldsymbol{\alpha})\\
&+f_{0}(\omega,\gamma_{1}(x),\boldsymbol{\alpha})\gamma_{1}(x)),\\
\end{aligned}
\end{equation}
where $\boldsymbol{\alpha}=(...\alpha_{-1},\alpha_{0},\alpha_{1}...)$ is the order parameter of the system and the functional of $\gamma_{1}(x)$ as
\begin{equation}
\alpha_{n}=\int(\gamma_{1}(x))^{n}q(x)dx, n=1,2,\dots.
\end{equation}
The solution of this integral differential equation Eq. (\ref{e15}) describes the structure of the local order parameter $\gamma_{1}(x)$ along $x$.

Up to now, we have discussed the system of all-connected phase oscillators in the limit of infinitely many oscillators and the case that only three Fourier coefficients of the coupling function are nonzero. The low-dimensional invariant manifold, namely the Poisson manifold, is got for both cases. We will see in the next two sections that these two conditions are exactly the scope of the OA ansatz. Two cases beyond this cope will be discussed respectively.

\section{Network with finite size}
We have considered the general model Eq. (\ref{e2}) in the limit of infinitely many oscillators, $N\rightarrow\infty$, in which we could get the phase density of oscillators $\rho(t,\varphi)$ and the corresponding continuity equation Eq. (\ref{ea1}). Order parameters $\alpha_{n}$ could be considered as Fourier coefficients of $\rho(t,\varphi)$, from which we get the equivalent expression of the dynamics of the system as the order parameter equations Eq. (\ref{e4}) and build the approach above.

In the limit $N\rightarrow\infty$, the macroscopic variable, namely the order parameter $\alpha(t)$, has the limit $\alpha(t)\rightarrow\alpha^{'}$ for steady states as the synchronous state and incoherent states, which gives us the basis to discuss the system analytically. In the case of a finite but large number of oscillators, i.e., $N\gg1$, the order parameter defined as $\alpha=(1/N)\sum_{j=1}^{N}e^{\mathrm{i}\varphi_{j}}$ will fluctuate around the value $\alpha^{'}$. This fluctuation depends on the number of oscillators as $O(N^{-\frac{1}{2}})$ \cite{Daido1986,Daido1989,Daido1990}. When the fluctuation is small enough, as $1/\sqrt{100000}\approx0.003$ for $N=100000$, the collective behaviors of the system with finite number of oscillators could be described by the approximation of infinitely many oscillators, for which analytical methods can be applied. Some analytical methods, e.g., the self-consistent method and the OA ansatz, are applicable for the case of infinitely many oscillators or equivalently the phase density description.

On the other hand, in the case of only a few number of oscillators, e.g., $N=10$, the difference $|\alpha(t)-\alpha^{'}|$ would be so large, as $1/\sqrt{10}\simeq0.316$, whose magnitude comparable with the value of $\alpha$. Obviously, it is not reliable to treat the system with the approximation $\alpha(t)\approx\alpha^{'}$ in this case. It is necessary to propose new approaches in analytically dealing with the collective behaviors of finite-oscillator systems.

\subsection{Ensemble approach}

In our approach, order parameters $\alpha_{n}$ could be considered as not only the Fourier coefficients of $\rho(t,\varphi)$ but also the collective variables as
\begin{equation}\label{ew7}
\alpha_{n}=\frac{1}{N}\sum_{j=1}^{N}e^{\mathrm{i}n\varphi_{j}},
\end{equation}
which does not depend on the approximation of infinitely many oscillators. For the system as
\begin{equation}
\dot\varphi_{j}(t)=F(\boldsymbol{\alpha},\varphi_{j},\boldsymbol{\beta}),j=1,\dots,N,
\end{equation}
with the definition of order parameters Eq. (\ref{ew7}), the dynamical equations for order parameters are
\begin{equation}\label{e16}
\dot\alpha_{n}=\frac{\mathrm{i}n}{K}\sum_{j=1}^{K}e^{\mathrm{i}n\varphi_{j}}\dot\varphi_{j},n=1,2,\dots.
\end{equation}
For the coupling function $F(\boldsymbol{\alpha},\varphi,\boldsymbol{\beta})$ is $2\pi$-periodic for $\varphi$, we have the Fourier expansion as
\begin{equation}
F(\boldsymbol{\alpha},\varphi,\boldsymbol{\beta})=\sum_{j=-\infty}^{\infty}f_{j}(\boldsymbol{\alpha},\boldsymbol{\beta})e^{\mathrm{i}j\varphi}.
\end{equation}
Substituting the expansion into Eq. (\ref{e16}), together with Eq. (\ref{ew7}), we have the closed form of equations for order parameters as
\begin{equation}\label{e17}
\dot\alpha_{n}=\mathrm{i}n\sum_{j=-\infty}^{\infty}f_{j}(\boldsymbol{\alpha},\boldsymbol{\beta})\alpha_{j+n},n=1,2,\dots,
\end{equation}
which is the same as we get in the limit of infinitely many oscillators.

In the case of a few number of oscillators, the definition of order parameters can also be considered as the coordinate transformation that transforms the microscopic variables $\{\varphi_{j}\}$ to macroscopic variables $\{\alpha_{n}\}$. Therefore it is inspiring that the dynamics of phase oscillators with corresponding initial conditions $\{\varphi_{j}(0)\}$ is equivalent to the dynamics of order parameters with corresponding initial conditions $\{\alpha_{n}(0)\}$. The number of oscillators, no matter finite or infinite, has no influence on the dynamical equations of order parameters. This forms the important basis of our approach.

Following this approach, when only the first three Fourier coefficients of the coupling function are nonzero, by setting $\alpha_{n}=\alpha_{1}^{n}$ and substituting the relation to the dynamical equations, the equations Eq. (\ref{e17}) for all the $\alpha_{n}$ will be reduced to a single one as
\begin{equation}
\dot\alpha_{1}=\mathrm{i}(f_{1}(\alpha_{1})\alpha_{1}^{2}+\bar f_{1}(\alpha_{1})+f_{0}(\alpha_{1})\alpha_{1}).
\end{equation}
It appears that we could get the Poisson manifold again even the size of the system is finite. However, the finite-size effect should be taken into account with care.

As a matter of fact, in the case of a finite number of oscillators, the Poisson manifold with $\alpha_{n}=\alpha_{1}^{n}$ is not attainable for the system. To see this, take the system of $N=2$ as an example. For the first two order parameters $\alpha_{1}$ and $\alpha_{2}$,  by definition
\begin{equation}\label{ew2}
\alpha_{1}=\frac{1}{2}\sum_{j=1}^{2}e^{\mathrm{i}\varphi_{j}}, \quad \quad \alpha_{2}=\frac{1}{2}\sum_{j=1}^{2}e^{2\mathrm{i}\varphi_{j}},
\end{equation}
if the system described by $\alpha_{1},\alpha_{2}$ is in the Poisson manifold, the relation $\alpha_{2}=\alpha_{1}^{2}$ could be rewritten in terms of the definition Eq. (\ref{ew7}) as
\begin{equation}\label{e18}
\frac{1}{2}\sum_{j=1}^{2}e^{2\mathrm{i}\varphi_{j}}=(\frac{1}{2}\sum_{j=1}^{2}e^{\mathrm{i}\varphi_{j}})^{2}.
\end{equation}
However, the equality Eq. (\ref{e18}) is not naturally valid. By using a simple calculation of the difference between the left-hand and the right-hand terms in Eq. (\ref{e18}) one obtains
\begin{equation}\label{e19}
\frac{1}{2}\sum_{j=1}^{2}e^{2\mathrm{i}\varphi_{j}}-(\frac{1}{2}\sum_{j=1}^{2}e^{\mathrm{i}\varphi_{j}})^2=\frac{1}{4}(e^{\mathrm{i}\varphi_{1}}-e^{\mathrm{i}\varphi_{2}})^{2}.
\end{equation}
This indicates that the system can evolve on the Poisson manifold only when $\varphi_{2}=\varphi_{1}$. In general, the relation $\alpha_{n}=\alpha_{1}^{n}$ gives exactly infinitely many independent constraints like Eq. (\ref{e18}) with $n=2,3,\dots$, and any solutions for finite oscillators will be determined as the trivial one as $\varphi_{j}=\varphi_{1}$ for all the index $j$, which means that except the synchronous state, all the states of the system of finite oscillators lie out of this Poisson manifold. The system can only evolve on the Poisson manifold in the limit of infinitely many oscillators.

Whereas, as a matter of fact, the equations Eqs. (\ref{e17}) hold for all $N$, whether the size of system is finite or infinite, and the synchronous state always satisfies the relation of the manifold. The Poisson manifold can be used at least approximately in the vicinity of synchronous state. However, we need some new methods and approximations.

Let us consider an ensemble with $M$ identical systems of $N$ oscillators which have the same dynamical equations and same parameters. The initial phases of oscillators are chosen from the same distribution, which makes the mean values of order parameters $\alpha_{n}$ over the ensemble have the limit when $M\rightarrow\infty$. This leads to the definition of the ensemble order parameter $\langle\alpha_{n}\rangle$ as
\begin{equation}
\langle\dot\alpha_{n}\rangle=\lim_{M\rightarrow\infty}\frac{1}{M}\sum_{j=1}^{M}\alpha_{n}^{(j)},n=1,2,\dots,
\end{equation}
where $M$ is the number of sampling systems in the ensemble and $\alpha_{n}^{(j)}$ is the $n$th order parameter for the $j$th system in the ensemble. Taking the ensemble average for both sides of the dynamical equations Eq. (\ref{e17}), we get
\begin{equation}\label{e20}
\langle\dot\alpha_{n}\rangle=\mathrm{i}n(\langle f_{1}\alpha_{n+1}\rangle+\langle\bar f_{1}\alpha_{n-1}\rangle+\langle f_{0}\alpha_{n}\rangle),
\end{equation}
where $n=1,2,\dots$, and $\langle\cdot\rangle$ means ensemble average.

In general Eq. (\ref{e20}) is not solvable because the terms in the right side as $\langle f_{1}\alpha_{n+1}\rangle$ cannot be simply described by $\langle\alpha_{n}\rangle$ in general. However, if $f_{j}$ is independent of all the systems in the ensemble as $\langle f_{j}\alpha_{n}\rangle=f_{j}\langle\alpha_{n}\rangle$, then we can get exactly the dynamical equations of the ensemble order parameters, as
\begin{equation}\label{e21}
\frac{d\langle\alpha_{n}\rangle}{dt}=\mathrm{i}n(f_{1}\langle\alpha_{n+1}\rangle+\bar f_{1}\langle\alpha_{n-1}\rangle+f_{0}\langle\alpha_{n}\rangle),
\end{equation}
where $n=1,2,\dots$ and the terms $f_{\pm1,0}$ may depend on $\langle\alpha_{n}\rangle$. Similar to Eq. (\ref{e17}), there is a solution for the ensemble order parameter equations Eqs. (\ref{e21}) as $\langle\alpha_{n}\rangle=\langle\alpha_{1}\rangle^{n}$, which is exactly the Poisson manifold.

Moreover, for more general cases $\langle f_{j}\alpha_{n}\rangle\neq f_{j}\langle\alpha_{n}\rangle$, but when the terms  $\langle f_{j}\alpha_{n}\rangle$ can be approximated by $\langle f_{j}\rangle\langle\alpha_{n}\rangle$, namely the statistical independence, we can also get the dynamical equations of the ensemble order parameters $\langle\alpha_{n}\rangle$ with similar approach. The validity of this approach can be measured by the error terms
\begin{equation}
e_{jn}=\langle f_{j}\alpha_{n}\rangle-\langle f_{j}\rangle\langle\alpha_{n}\rangle,
\end{equation}
with $j=\pm1,0$ and $n=1,2,\dots$.

\subsection{The star Sakaguchi-Kuramoto model}

In the following, let us take the star Sakaguchi-Kuramoto model as an example, which is a typical topology and model in grasping the essential properties of heterogeneous networks and synchronization process, as
\begin{equation}
\begin{aligned}
\dot{\theta}_{h} &=\omega_{h}+\lambda\sum_{j=1}^{N}\sin(\theta_{j}-\theta_{h}-\beta),\\
\dot{\theta}_{j} &=\omega+\lambda\sin(\theta_{h}-\theta_{j}-\beta),
\end{aligned}\end{equation}
where $1\leq j\leq N$, $\omega_{h},\theta_{h}$ and $\omega_{j},\theta_{j}$ are the natural frequency and the phase of hub and leaf nodes respectively, $\lambda$ is the coupling strength and $\beta$ is the phase shift. By introducing the phase difference $\varphi_{j}=\theta_{h}-\theta_{j}$, the dynamical equation can be transformed into
\begin{equation}\label{ea2}
\dot\varphi_{j}=\Delta\omega-\lambda\sum_{k=1}^{N}\sin(\varphi_{k}+\beta)-\lambda\sin(\varphi_{j}-\beta),
\end{equation}
where $\Delta\omega=\omega_{h}-\omega_{j}$ is the natural-frequency difference between hub and leaf nodes.

By introducing the order parameter $re^{\mathrm{i}\Phi}\equiv\alpha=\frac{1}{N}\sum_{j=1}^{N}e^{\mathrm{i}\varphi_{j}}$, we could rewrite the dynamics as
\begin{equation}
\dot\varphi_{j}=fe^{\mathrm{i}\varphi_{j}}+g+\bar{f}e^{-\mathrm{i}\varphi_{j}},\quad j=1,\cdots,N,
\end{equation}
where $f=\mathrm{i}\frac{\lambda}{2}e^{-\mathrm{i}\beta}$, $g=\Delta\omega-\lambda Nr\sin(\Phi+\beta)$. Hence the system is in the framework which we discussed above, with the first three nonzero Fourier coefficients of the coupling function. With our approach, the dynamical equations for order parameters can be obtained as
\begin{equation}\label{e23}
\dot\alpha_{n}=\mathrm{i}n(f\alpha_{n+1}+\bar f\alpha_{n-1}+g\alpha_{n}),n=1,2,\dots.
\end{equation}

In this specific model, we cannot simply make the approximation as $N\rightarrow\infty$ because $g=\Delta\omega-\lambda Nr\sin(\Phi+\beta)$ diverges with $N\rightarrow\infty$. There are two ways in further simplifying the system. First, we could set a hypothetical system as
\begin{equation}\label{e24}
\dot\vartheta_{j}=\Delta\omega-N\lambda\sum_{j=1}^{M}\frac{1}{M}\sin(\vartheta_{j}+\beta)-\lambda\sin(\vartheta_{j}-\beta),
\end{equation}
where $1\leq j\leq M$, $\vartheta_{j}$ is the hypothetical phase oscillator and $N$ is considered as a parameter in this system. Defining the order parameter for this system as $re^{\mathrm{i}\Phi}\equiv\alpha=\frac{1}{M}\sum_{j=1}^{M}e^{\mathrm{i}\vartheta_{j}}$, Eq. (\ref{e24}) becomes
\begin{equation}
\dot\varphi_{j}=fe^{\mathrm{i}\varphi_{j}}+g+\bar{f}e^{-\mathrm{i}\varphi_{j}},\quad j=1,\cdots,M,
\end{equation}
where $f=\mathrm{i}\frac{\lambda}{2}e^{-\mathrm{i}\beta}$, $g=\Delta\omega-\lambda Nr\sin(\Phi+\beta)$. Following our approach, the corresponding order parameter equations read
\begin{equation}\label{e25}
\dot\alpha_{n}=\mathrm{i}n(f\alpha_{n+1}+\bar f\alpha_{n-1}+g\alpha_{n}),n=1,2,\dots,
\end{equation}
which are exactly the same as Eq. (\ref{e23}). But for this hypothetical system, we could take the limit of infinitely many oscillators as $M\rightarrow\infty$, which could be discussed analytically with the traditional OA anstaz.

Obviously, these two systems are quite different for phase variables, one with finite oscillators the other with infinitely many oscillators, but they have the same order parameter equations. In the following we will see that the hypothetical system Eq. (\ref{e24}) is exactly a representation of the ensemble of the original model Eq. (\ref{ea2}).

On the other hand, following our ensemble approach, for the system Eq. (\ref{ea2}), let us choose an ensemble consisting of systems with the same parameters $\lambda, \Delta\omega$ and $N$, and different initial conditions chosen from the same distribution, as the Poisson kernel. We have the ensemble average of the dynamical equation as
\begin{equation}\label{e26}
\begin{aligned}
\langle\dot\alpha_{n}\rangle&=\mathrm{i}n(\mathrm{i}\frac{\lambda}{2}e^{-\mathrm{i}\beta}\langle\alpha_{n+1}\rangle-\mathrm{i}\frac{\lambda}{2}e^{\mathrm{i}\beta}\langle\alpha_{n-1}\rangle+\Delta\omega\langle\alpha_{n}\rangle\\
                &+\frac{\mathrm{i}}{2}\lambda N(\langle\alpha_{1}\alpha_{n}\rangle e^{\mathrm{i}\beta}-\langle\bar\alpha_{1}\alpha_{n}\rangle e^{-\mathrm{i}\beta})),
\end{aligned}
\end{equation}
where $n=1,2,\dots$. Setting
\begin{equation}
\begin{aligned}
\langle\alpha_{1}\alpha_{n}\rangle=\langle\alpha_{1}\rangle\langle\alpha_{n}\rangle+e_{n1},\\
\langle\bar\alpha_{1}\alpha_{n}\rangle=\langle\bar\alpha_{1}\rangle\langle\alpha_{n}\rangle+e_{n2},
\end{aligned}
\end{equation}
for $n=1,2,\dots$. If $|e_{n1}|\ll|\langle\alpha_{1}\rangle\langle\alpha_{n}\rangle|$ and $|e_{n2}|\ll|\langle\bar\alpha_{1}\rangle\langle\alpha_{n}\rangle|$, or typically $|e_{n1}|\ll1$ and $|e_{n2}|\ll1$, they can be seen as the perturbation terms for the dynamics of ensemble order parameters in Eq. (\ref{e26}). Ignoring the perturbations, we could get
\begin{equation}\label{e27}
\begin{aligned}
\langle\dot\alpha_{n}\rangle&=\mathrm{i}n(\mathrm{i}\frac{\lambda}{2}e^{-\mathrm{i}\beta}\langle\alpha_{n+1}\rangle-\mathrm{i}\frac{\lambda}{2}e^{\mathrm{i}\beta}\langle\alpha_{n-1}\rangle\\
                &\Delta\omega\langle\alpha_{n}\rangle+\frac{\mathrm{i}}{2}\lambda N(\langle\alpha_{1}\rangle\langle\alpha_{n}\rangle e^{\mathrm{i}\beta}\\
                &-\langle\bar\alpha_{1}\rangle\langle\alpha_{n}\rangle e^{-\mathrm{i}\beta})),
\end{aligned}
\end{equation}
and this is the same as Eq. (\ref{e25}) for hypothetical system. In this case, the ensemble order parameters are the same as the order parameters for the hypothetical system.

For the case of a finite but large number of oscillators, the hypothetical system is exactly the continuous model which shares the same order parameter equations but with infinitely many oscillators. As in the case of a few oscillators, the hypothetical system is a presentation of the ensemble for the systems, where the difference terms $e_{n1}$ and $e_{n2}$ by this method will not only introduce some fluctuations but also some systematic errors, which can be described by
\begin{equation}
\begin{aligned}
E_{n}&=|\langle\alpha_{1}\alpha_{n}\rangle-\langle\alpha_{1}\rangle\langle\alpha_{n}\rangle|,\\
E_{n}^{'}&=|\langle\bar\alpha_{1}\alpha_{n}\rangle-\langle\bar\alpha_{1}\rangle\langle\alpha_{n}\rangle|,
\end{aligned}
\end{equation}
with $n=1,2,\dots$ for this specific model.

In terms of numerical computation, we can check the above approximation by examining $E_{n}$ and $E_{n}^{'}$ for the ensemble of $M$ systems. The approximation $E_{n}$ and $E_{n}^{'}$ can be divided into two parts as $E_{n}=f_{n}+s_{n}$ and $E_{n}^{'}=f_{n}^{'}+s_{n}^{'}$ where $f_{n}$ and $f_{n}^{'}$ are the fluctuation parts proportional to $1/\sqrt{M}$ which depend on the size of the ensemble and  $s_{n}$ and $s_{n}^{'}$ are the systematic parts which are introduced by the statistical independence assumption.

\begin{figure}
\includegraphics[height=7cm]{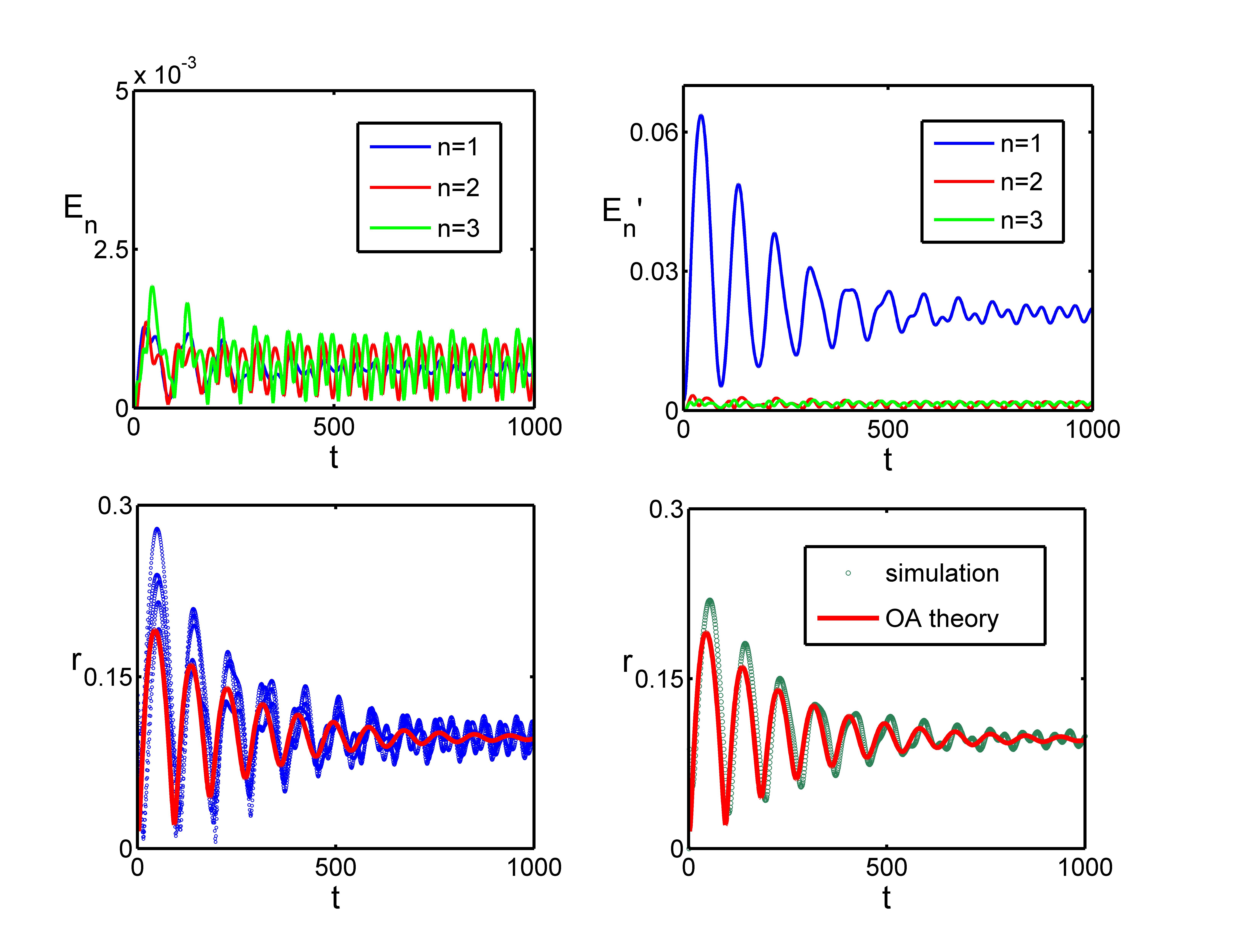}
\caption{(color online) (a) and (b) The first three error terms $E_{n}$ and $E_{n}^{'}$ versus time for the system with $N=10$. (c) and (d) Order parameters getting from numerical simulation and approximated OA ansatz. Results from approximated OA ansatz is the red line, and every single simulation results is the blue lines in (c) with ensemble average the light blue line in (d).} \label{fig:2}
\end{figure}

Take $N=10$ as an example. When $M=1000$ the fluctuation parts are ignorable. We find that except for $E_{1}^{'}$ is around $0.03$, all of $E_{n}$ and $E_{n}^{'}$ for $n\leq 3$ are much smaller than $0.01$ in this case as plotted in Fig. \ref{fig:2}(a) and (b). Comparing the typical magnitude of values of order parameters as $10^{-1}$, the approximation of the ensemble approach is obviously reasonable.

For the initial conditions chosen from the Poisson kernel, with the dynamical equations Eq. (\ref{e27}), the ensemble order parameters will evolve on an invariant manifold, i.e., the Poisson manifold with $\langle\alpha_{n}\rangle=\langle\alpha_{1}\rangle^{n}$. Denote $\langle\alpha_{1}\rangle$ by $z$, we have
\begin{equation}\label{ea5}
\dot z=-\dfrac{\lambda}{2}z^{2}e^{-\mathrm{i}\beta}+\mathrm{i}(\Delta \omega+\frac{\mathrm{i}}{2}\lambda N(ze^{\mathrm{i}\beta}-\bar z e^{-\mathrm{i}\beta}))z+\dfrac{\lambda}{2}e^{\mathrm{i}\beta}.
\end{equation}
The difference between $z$ and the ensemble average of numerical simulations is checked in Fig. \ref{fig:2}(d), which shows that it is reasonable to describe the behavior of the ensemble by $z$. Moreover, for every single system in the ensemble, with some fluctuation, it can also be described by the ensemble order parameter and hence by $z$, as shown in Fig. \ref{fig:2}(c).

With this ensemble assumption works, the two-dimensional equation Eq. (\ref{ea5}) in the bounded space as $|z|\leq 1$ describes the dynamics of the coupled oscillator system. Every stationary state for phase oscillators system has its counterpart in this space for $z=re^{\mathrm{i}\Phi}$. For example, the in-phase state(IPS) defined as $\varphi_{j}(t)=\varphi_{q}(t)=\varphi_{t}$ corresponds to a limit cycle as $r(t)=1$. The synchronous state (SS) defined as $\varphi_{j}(t)=\varphi_{q}(t)=\varphi$ corresponds to a fixed point with $r(t)=1$ and $\Phi(t)=c$. The splay state(SPS) defined as  $\varphi_{j}(t)=\varphi(t+\frac{jT}{K})$ with $T$ the period of $\varphi(t)$ corresponds to a fixed point with $r(t)=r_{0}<1$ and $\Phi(t)=c$. In the two-dimensional order-parameter space, it is easy to analyze the existence and stability conditions of these states, and the basion of attraction for each state in the coexistence region can also be conveniently determined \cite{xu}. This is the reason why we introduce the invariant manifold governed by low-dimensional dynamics.

As we see, in the case with only finite oscillators, the ensemble order parameter plays the dominant role in revealing the key collective structure of the system. The price of the ensemble order parameter description is the error induced by the statistical independence assumption, whose validity can be checked by numerical results. In the next section we will see that when more than three Fourier coefficients of the coupling function are nonzero, where the traditional order parameter scheme fails, one can still discuss the collective dynamics in terms of the ensemble order parameter approach.

\section{Complex coupling function}
In the above sections, for the coupling functions with only first three terms as
\begin{equation}\label{ea3}
F(\boldsymbol{\alpha},\varphi)=f_{1}(\boldsymbol{\alpha})e^{\mathrm{i}\varphi}+f_{-1}(\boldsymbol{\alpha})e^{-\mathrm{i}\varphi}+f_{0}(\boldsymbol{\alpha}),
\end{equation}
we have the dynamical equations for order parameters $\alpha_{n}$ as
\begin{equation}\label{e28}
\dot\alpha_{n}=\mathrm{i}n(f_{1}\alpha_{n+1}+\bar f_{1}\alpha_{n-1}+f_{0}\alpha_{n}),n=1,2,\dots.
\end{equation}
On the invariant manifold $\alpha_{n}=\alpha_{1}^{n}$, all these equations are reduced to a single one
\begin{equation}\label{ea4}
\dot\alpha_{1}=\mathrm{i}(f_{1}(\alpha_{1})\alpha_{1}^{2}+\bar f_{1}(\alpha_{1})+f_{0}(\alpha_{1})\alpha_{1}).
\end{equation}
For general situations, the coupling function can not be simply truncated to only the first terms. It is thus important to consider the order parameter approach for more complicated cases.

\subsection{Higher order coupling}

Firstly, let us consider a coupling function with higher order Fourier coefficients like
\begin{equation}\label{e29}
G(\boldsymbol{\alpha^{\prime}},\phi)=g_{m}(\boldsymbol{\alpha^{\prime}})e^{\mathrm{i}m\phi}+g_{-m}(\boldsymbol{\alpha^{\prime}})e^{-\mathrm{i}m\phi}+g_{0}(\boldsymbol{\alpha^{\prime}}),
\end{equation}
where $m>1$ is an integer and $\boldsymbol{\alpha^{\prime}}$ are the order parameters for the system. Note that Eq. (\ref{e29}) can be regarded as the transformation of Eq. (\ref{ea3}) with $\varphi=m\phi$ and $f_{j}=g_{j\cdot m}$. The order parameters $\boldsymbol{\alpha}$ for $\varphi$ is related to the order parameters $\boldsymbol{\alpha^{\prime}}$ for $\phi$ as
\begin{equation}
\alpha_{n}=\frac{1}{K}\sum_{j=1}^{K}e^{\mathrm{i}n\varphi_{j}}=\frac{1}{K}\sum_{j=1}^{K}e^{\mathrm{i}n\cdot m\phi_{j}}=\alpha_{n\cdot m}^{\prime}.
\end{equation}
And the phase density of $\phi$ with the transformation reads
\begin{equation}
\begin{aligned}
\rho^{\prime}(\phi)&=\rho(\varphi)\\
&=\frac{1}{2\pi}\sum_{j=-\infty}^{\infty}\alpha_{j}e^{\mathrm{i}j\varphi}=\frac{1}{2\pi}\sum_{j=-\infty}^{\infty}\alpha_{j\cdot m}^{\prime}e^{\mathrm{i}j\cdot m\phi}.
\end{aligned}
\end{equation}
This indicates that all the terms $\alpha_{n}^{\prime}$ with $|n|\neq j\cdot m, j=0,1,2...$ are zero. Moreover, we can transform the Poisson manifold $\alpha_{n}=(\alpha_{1})^{n}$ for $\varphi$ to the invariant manifold $\alpha_{n\cdot m}^{\prime}=(\alpha_{m}^{\prime})^{n}$ for $\phi$, which is governed by
\begin{equation}
\dot\alpha_{m}^{\prime}=\mathrm{i}m(f_{m}^{\prime}(\alpha_{m}^{\prime})^{2}+\bar f_{m}^{\prime}+f_{0}^{\prime}\alpha_{m}^{\prime}),
\end{equation}
corresponding to Eq. (\ref{ea4}) for $\alpha_{1}$. This is exactly the low-dimensional manifold for the coupling function Eq. (\ref{e29}) with higher order Fourier coefficients. This manifold is based on the $m$th order parameter $\alpha_{m}^{\prime}$ and the statement that all the terms $\alpha_{n}^{\prime}$ with $|n|\neq j\cdot m, j=0,1,2...$ are zero. This is a bit queer in this respect.

Let us take the case $m=2$ as an example \cite{Skardal2011higer}. In this case only $f_{\pm2}$ and $f_{0}$ among the Fourier components of the coupling function are nonzero, i.e.,
\begin{equation}
F(\boldsymbol{\alpha},\varphi)=f_{2}(\boldsymbol{\alpha})e^{2\mathrm{i}\varphi}+f_{-2}(\boldsymbol{\alpha})e^{-2\mathrm{i}\varphi}+f_{0}(\boldsymbol{\alpha}).
\end{equation}
According to the above analysis, the system has the invariant manifold as $\alpha_{2n+1}=0,\alpha_{2n+2}=\alpha_{2}^{n+1},n\geq 0$, or equivalently $\alpha_{n}=1/2\alpha_{2}^{n/2}[(-1)^{n}+1]$, with dynamics
\begin{equation}
\dot\alpha_{2}=2\mathrm{i}(f_{2}\alpha_{2}^{2}+\bar f_{2}+f_{0}\alpha_{2}).
\end{equation}
The phase density of oscillators in this case is
\begin{equation}\label{e30}
\rho(\varphi)=\frac{1}{2\pi}\sum_{j=-\infty}^{\infty}\alpha_{2\cdot j}e^{2\mathrm{i}j\phi}.
\end{equation}

By using our order-parameter-analysis approach, for the case of $m=2$, it is not hard to get the the equations of order parameters $\alpha_{n}$ as
\begin{equation}\label{e31}
\begin{aligned}
\dot\alpha_{2n+1}&=\mathrm{i}(2n+1)(f_{2}\alpha_{2n+3}+\bar f_{2}\alpha_{2n-1}+f_{0}\alpha_{2n+1}),\\
\dot\alpha_{2n}&=\mathrm{i}(2n)(f_{2}\alpha_{2n+2}+\bar f_{2}\alpha_{2n-2}+f_{0}\alpha_{2n}),
\end{aligned}
\end{equation}
with $n=1,2,\dots$. The manifold with $\alpha_{2n+1}=0,\alpha_{2n}=\alpha_{2}^{n},n\geq 0$ is obviously the solution of these equations. This invariant manifold separates the relation $\alpha_{n}=\alpha_{1}^{n}$ into two parts, where the part $\alpha_{2n}=\alpha_{2}^{n}$ conserves the relation, and the odd part $\alpha_{2n+1}=0$ can be regarded as a special choice of the relation. Even if in the state defined as $|\alpha_{2n}|=|\alpha_{2}|^{n}=1$, we still have $\alpha_{1}=0$, which represents the state of cluster synchrony as discussed in \cite{Skardal2011higer}. And the manifold with $\alpha_{2n+1}=0,\alpha_{2n}=\alpha_{2}^{n},n\geq 0$ can be regarded as a sub-manifold of the Poisson manifold.

An interesting issue here is the reason why we set all the $\alpha_{2n+1}=0$ other than $\alpha_{2n+1}=\alpha_{1}^{2n+1}$. To see this, we just substitute the relation $\alpha_{2n+1}=\alpha_{1}^{2n+1},n\geq0$ into the odd part of Eq. (\ref{e31}), then the equations will be reduced to the following equations,
\begin{equation}\label{ew4}
\begin{aligned}
\dot\alpha_{1}&=\mathrm{i}(f_{2}\alpha_{1}^{3}+\bar f_{2}\bar\alpha_{1}+f_{0}\alpha_{1}),\\
\dot\alpha_{1}&=\mathrm{i}(f_{2}\alpha_{1}^{3}+\bar f_{2}\alpha_{1}^{-1}+f_{0}\alpha_{1}),
\end{aligned}
\end{equation}
where the first equation is got from $n=0$ in Eq. (\ref{e31}), and the second equation is got from all the others $n>0$ in Eq. (\ref{e31}). Comparing the two equations in Eq. (\ref{ew4}), the consistent condition requires  $\bar\alpha_{1}=\alpha_{1}^{-1}$ , and this implies that $|\alpha_{1}|=1$. Hence, there isn't a general relation of $\alpha_{2n+1}$, apart from the vicinity of synchronous state. Thus we should consider another special solution for the odd part of Eq. (\ref{e31}) as $\alpha_{2n+1}=0,n\geq 0$.

On the other hand, for the even part of Eq. (\ref{e31}), by setting $\alpha_{2n}=\alpha_{2}^{n}$, the equations will be reduced to a single one as
\begin{equation}\label{ew5}
\dot\alpha_{2}=2i(f_{2}\alpha_{2}^{2}+\bar f_{2}+f_{0}\alpha_{2}).
\end{equation}
The conjugate part $\alpha_{-2}$ is separated from $\alpha_{2}$ naturally in this equation, and we can get the low-dimensional manifold governed by the order parameter equation Eq. (\ref{ew5}), as a sub-manifold of the Poisson manifold.

\subsection{The Dominating-term assumption}

As mentioned above, the Poisson manifold no longer exists when the order parameter relation $\alpha_{n}=\alpha_{1}^{n}$ can not be used to reduce the order parameter equations to the single one. For the case of only higher terms this problem can be fixed by separating the order parameters into groups and setting some of them zero with which a special sub-manifold of the Poisson manifold can be used to get low-dimensional behaviors. However, when more than three terms in Fourier expansion are considered, even the sub-manifold on longer exists.

Consider a coupling function with the first five nonzero Fourier coefficients, as
\begin{equation}
F(\boldsymbol{\alpha},\varphi)=\sum_{n=-2}^{2}f_{n}(\boldsymbol{\alpha})e^{\mathrm{i}n\varphi},
\end{equation}
the corresponding equations of order parameters read now
\begin{equation}\label{e32}
\dot\alpha_{n}=\mathrm{i}n(f_{2}\alpha_{n+2}+\bar f_{2}\alpha_{n-2}+f_{1}\alpha_{n+1}+\bar f_{1}\alpha_{n-1}+f_{0}\alpha_{n}),
\end{equation}
with $\alpha_{-n}=\bar\alpha_{n}$, $n=1,2,\dots$. In this case, if we suppose $\alpha_{n}=\alpha_{1}^{n}$, then two different dynamical equations will be obtained
\begin{equation}
\begin{aligned}
\dot\alpha_{1}&=\mathrm{i}(f_{2}\alpha_{1}^{3}+\bar f_{2}\bar\alpha{1}+f_{1}\alpha_{1}^{2}+\bar f_{1}+f_{0}\alpha_{1}),\\
\dot\alpha_{1}&=\mathrm{i}(f_{2}\alpha_{1}^{3}+\bar f_{2}\alpha_{1}^{-1}+f_{1}\alpha_{1}^{2}+\bar f_{1}+f_{0}\alpha_{1}).
\end{aligned}
\end{equation}
Hence we will have $\alpha_{-1}=\alpha_{1}^{-1}$ as the requirement of coincidence of these two equations. The relation of $\alpha_{n}=\alpha_{1}^{n},n>0$ is broken by the conjugate terms and there is no way in finding out a separated solution as we did for Eq. (\ref{e31}). The only solutions with the relation $\alpha_{n}=\alpha_{1}^{n}$ for this system seems to be either $\alpha_{n}=\alpha_{1}^{n}=0$ or $|\alpha_{n}|=|\alpha_{1}|^{n}=1$. None of them are expected for our analysis, unless in the vicinity of synchronous state with $|\alpha_{1}|=1$ or in the vicinity of the incoherent state with $\alpha_{1}=0$.

On the other hand, for $n\geq2$ all the dynamical equations for the $n$th order parameter in Eq. (\ref{e32}) are reduced to the same equation by $\alpha_{n}=\alpha_{1}^{n}$, and the symmetry of Eq. (\ref{e32}) can be represented by $\alpha_{n}=\alpha_{1}^{n}$ at least for all the order parameter equations for $n\geq2$. Moreover, due to the bound $|\alpha_{1}|\leq1$, the higher the order components $\alpha_{n}$ become smaller with increasing $n$. If we choose the initial conditions as $\alpha_{n}=\alpha_{1}^{n}$, then along the evolution of Eq. (\ref{e32}), the differences of higher terms $\alpha_{n}$ with $\alpha_{1}^{n}$ will keep small enough. By ignoring the differences and making the approximation that $\alpha_{n}\approx\alpha_{1}^{n}$, we can obtain the dynamical equations for $\alpha_{1}$ as
\begin{equation}
\dot\alpha_{1}=\mathrm{i}(f_{2}\alpha_{1}^{3}+\bar f_{2}\bar\alpha{1}+f_{1}\alpha_{1}^{2}+\bar f_{1}+f_{0}\alpha_{1}),
\end{equation}
which can be regarded as the main term of the dynamics Eq. (\ref{e32}). From now on we call this approach \textit{dominating-term assumption}. We will further show that this approximation is pretty efficient in dealing with systems with complicated coupling functions.

Let us take the system described by Eq. (\ref{ea2}) as an example. By considering higher terms, we have the dynamical equations
\begin{equation}
\dot\varphi_{j}=\mathrm{i}(f_{1}e^{\mathrm{i}\varphi_{j}}+g+\bar{f_{1}}e^{-\mathrm{i}\varphi_{j}}+f_{2}e^{2\mathrm{i}\varphi_{j}}+\bar{f_{2}}e^{-2\mathrm{i}\varphi_{j}}),
\end{equation}
where $j=1,\cdots,N$ and $f_{1}=\mathrm{i}\frac{\lambda_{1}}{2}$, $f_{2}=\mathrm{i}\frac{\lambda_{2}}{2}$, $g=\Delta\omega-\lambda Nr\sin(\Phi)$ with $re^{\mathrm{i}\Phi}=\alpha_{1}$. By introducing the dominating-term assumption, the dynamical equation for $\alpha_{1}$ reads
\begin{equation}\label{e33}
\dot\alpha_{1}=\mathrm{i}(f_{2}\alpha_{1}^{3}+\bar f_{2}\bar\alpha_{1}+f_{1}\alpha_{1}^{2}+\bar f_{1}+g\alpha_{1}).
\end{equation}
If the dominating-term assumption works, Eq. (\ref{e33}) can be used to study the collective behaviors of the coupled oscillators, which can be regarded as a sort of approximate low-dimensional behaviors.

\begin{figure}
\includegraphics[height=3.5cm]{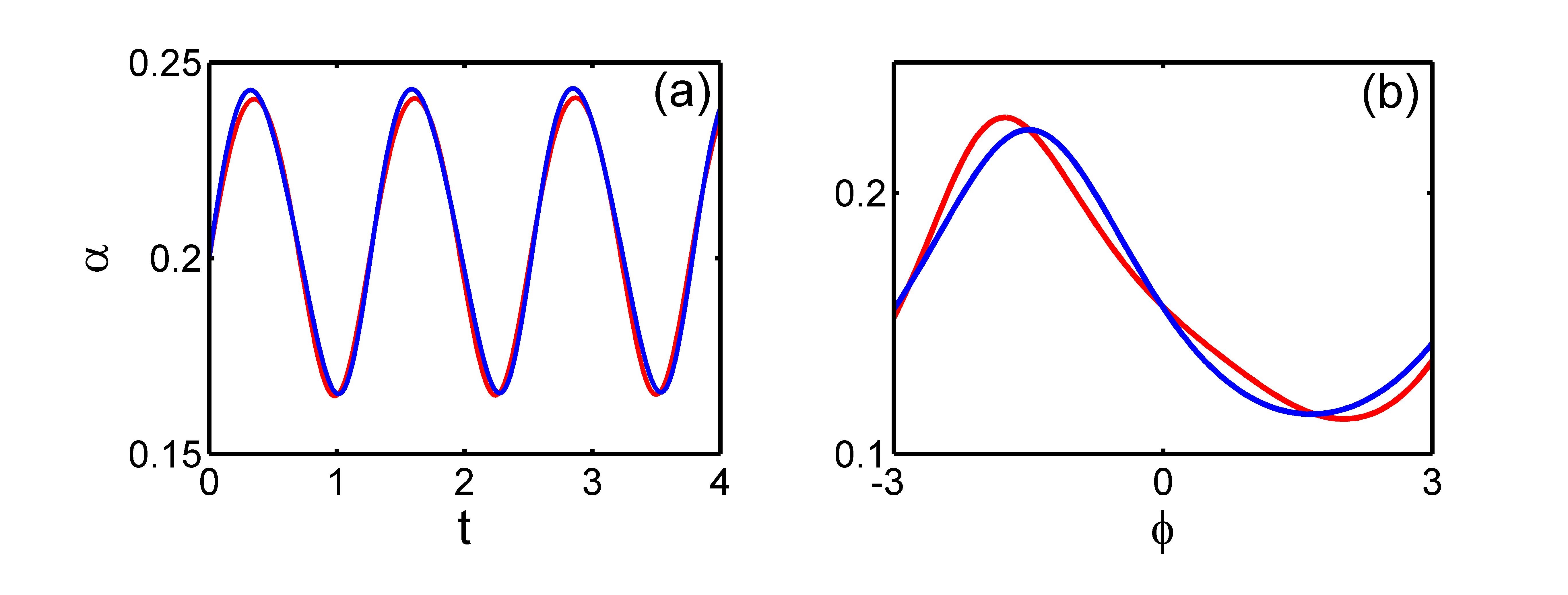}
\caption{(color online)(a) Order parameters getting from numerical simulation (red line) and approximate OA ansatz Eq.(\ref{e33})(blue line). (b) Distribution of phase variables getting form numerical simulation (red) and reconstituting Poisson kernel with order parameter $\alpha_{1}$ (blue). The system considered as $\lambda=\lambda_{1}=0.4,\lambda_{2}=0.3,N=1,\Delta\omega=5$} \label{fig:5}
\end{figure}

The validity of the dominating-term assumption can be checked via numerical simulations. For the initial state chosen from the Poisson kernel distribution, by using numerical simulation, we have the order parameter from the continuity equation of the phase density and the approximate manifold. In Fig. \ref{fig:5}(a) the order parameter $\alpha_{1}$ is plotted. It can be found that the result in terms of the dominating-term approximation coincides very well with numerical results. In Fig. \ref{fig:5}(b), the density of oscillators at a given time is also plotted, and the result shows that though the distribution is indeed no longer the Poisson kernel distribution, it is instead close to it. This gives us the mechanism of how this assumption works.

According to numerical results, we can see that the dominating-term assumption works pretty well, and the system shows approximated low-dimensional behaviors. With the dynamical equations for the first order parameter Eq. (\ref{e33}), we can do the study further to find the property of collective behaviors of the system, which will be discussed in detail in another paper.

\section{Discussion}

In this paper, we focused on the theoretical description of low-dimensional order-parameter dynamics of the collective behaviors in coupled phase oscillators. We have derived the closed form of dynamical equations for order parameters, from which we find the Poisson kernel as an invariant manifold and the well-known OA ansatz. Different from the traditional Ott-Antonsen ansatz and Watanabe-Strogatz's approach, our approach is suitable for systems of both finite and infinite, identical and nonidentical oscillators. In our approach, the scope of the OA ansatz is determined as two parts, i.e.,  the limit of infinitely many oscillators and the condition that only the first three Fourier coefficients of the coupling strength are nonzero.

By using our order parameter analysis, we also discussed two cases that go beyond the scope of the OA ansatz, i.e., the case of a finite-size system and the case of coupled oscillator systems with more complicated coupling functions. We have discussed the reasons why the OA ansatz cannot be used directly to these two cases and we further developed the approximation methods to deal with these difficulties with the OA ansatz. We developed two methods, namely the ensemble method and dominating-term assumption. It is shown that these schemes work pretty well, and their validity has been checked by the numerical simulations.

For the case of a finite number of coupled oscillators, it is shown that the system is out of the domain of low-dimensional manifold of the coupled order parameter equations. Hence, from the view of the geometry theory of manifolds, the finite size of system brings some fluctuations along the manifold which is introduced from initial states. If the dynamics of order parameter equations is not too complicated and the manifold satisfying $\alpha_{n}=\alpha_{1}^{n}$ can be used to describe the dynamics approximately, all the trajectories from different initial conditions around the Poisson manifold will be described by the mechanism of this manifold approximately. This is the meaning of the ensemble method and the reason for why it works in some models.

In other case, with a more complicated coupling function, the Poisson manifold is not a solution of the dynamical system. However, for the infinitely many coupled equations for $\alpha_{n}$, except for few of them, the relation $\alpha_{n}=\alpha_{1}^{n}$ indeed reflects the symmetry of these order parameter equations. When we choose the initial states in this manifold, the evolution of the dynamics will certainly be influenced by the symmetry of it, which is described by $\alpha_{n}=\alpha_{1}^{n}$ approximately. The method with dominating-term assumption is indeed designed to show this effect, and consequently the system is simplified by it for some models.

In this paper, we have developed the order parameter analysis, with which we get the low-dimensional behaviors of coupled phase oscillators, such as the OA ansatz, the ensemble order parameter approach and the dominating-term approximation. However, to fully understand the low-dimensional collective behaviors we need a more comprehensive understanding of the Poisson manifold and its relation with the dynamics of the system, which is still an open question. We believe that the order parameter analysis should be a powerful tool in helping to reveal the mechanism of low-dimensional collective behaviors.

\section{Acknowledgment}

This work is partially supported by the National Natural Science Foundation of China (Grant No. 11075016 and 11475022).

\end{document}